**Université de Jendouba**

**I**nstitut **S**upérieur d'**I**nformatique de **K**ef

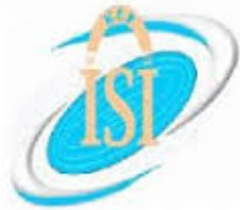

**RAPPORT DE MÉMOIRE DE FIN D'ÉTUDES**

Présenté en vue de l'obtention du

**DIPLÔME DE MASTÈRE DE RECHERCHE**

**EN SYSTÈME D'INFORMATION ET WEB**

**Sujet : L'alignement des réseaux d'Interactions Protéine-Protéine.**

Réalisé par

*Mlle Ghanjeti Sarra*

**Soutenu le : 13/06/2019 devant le jury composé de :**

Président : Dr. Cherni Ibtissem (Maitre-Assistante, ISI- Kef, Université Jendouba)

Rapporteur : Dr. Louati Aymen(Maitre-Assistant, ISI- Kef, Université Jendouba)

Directeur du mémoire : Dr. Warith Eddine DJEDDI (Maitre-Assistant, ISSAT-Kasserine, Université Kairouan)-(LIPAH)

**Année universitaire : 2017-2018**

# Liste des abréviations

| | |
|---|---|
| **IPP** | **I**nteraction **P**rotéine-**P**rotéine |
| **DIP** | **D**atabase of  **I**nteracting **P**roteins |
| **MINT** | **M**olecular **INT**eraction database |
| **HPRD** | **H**uman **P**rotein **R**eference **D**atabase |
| **STRING** | **S**earch **T**ool for the **R**etrieval of **I**nteracting **G**enes/Proteins |
| **BioGRID** | **Bio**logical General **R**epository for **I**nteraction **D**atasets |
| **IsoBase** | **Iso**Rank PPI Network Alignment **Base**d Ortholog Database |
| **BP** | **B**iological **P**rocess |
| **MF** | **M**olecular **F**unction |
| **CC** | **C**ellular **C**omponent |
| **DAG** | **D**irected **A**cyclic **G**raphs |
| **BLAST** | **B**asic **L**ocal **A**lignment **S**erach **T**ool |
| **EC** | **E**dges **C**orrectness |
| **LCCS** | **L**argest **C**ommon **C**onnected **S**ubgraph |
| **MNE** | **M**ean **N**ormalized **E**ntropy |
| **ME** | **M**ean **E**ntropy |
| **PPINA** | **P**rotein **P**rotein **I**ntercation **N**etwork **A**ligner |

# Résumé

L'alignement de réseaux d'IPPs, vise à trouver des similarités topologiques et fonctionnelles entre les réseaux de différentes espèces.

Plusieurs approches d'alignements de réseaux d'IPPs ont été proposées. Chacune de ces approches repose sur une méthode d'alignement différente et utilise différentes informations biologiques pendant le processus d'alignement tel que la structure topologique de réseaux et les similarités séquentielles entre les protéines, mais moins d'entre eux intègrent les similarités fonctionnelles entre les protéines.

Dans ce cadre, nous présentons notre algorithme PPINA (Protein-Protein Interaction Network Aligner), une extension de l'algorithme NETAL. Ce dernier aligne deux réseaux d'IPPs en se basant sur la similarité séquentielle, fonctionnelle et topologique des protéines.

PPINA a été testé sur de réseaux d'IPPs réels. Les résultats montrent que PPINA a surpassé d'autres algorithmes d'alignement où il fournit des résultats biologiquement significatifs.

**MOT-CLES :**

L'alignement de réseaux d'IPPs, le processus d'alignement, la structure topologique de réseaux, les similarités séquentielles, les similarités fonctionnelles, PPINA.

# Abstract

PPI network alignment, aims to find topological and functional similarities between networks of different species.

Several alignment approaches have been proposed. Each of these approaches relies on a different alignment method and uses different biological information during the alignment process such as the topological structure of the networks and the sequence similarities between the proteins, but less of them integrate the functional similarities between proteins.

In this context, we present our algorithm PPINA (Protein-Protein Interaction Network Aligner), which is an extension of the NETAL algorithm. The latter aligns two networks based on the sequence, functional and network topology similarity of the proteins.

PPINA has been tested on real PPI networks. The results show that PPINA has outperformed other alignment algorithms where it provides biologically meaningful results.

# Table des matières

# Liste des figures

# Liste des tableaux

# Introduction générale

La plupart des protéines réalisent leurs fonctions moléculaires en interagissant avec d'autres protéines, forment un réseau d'interaction Protéines Protéines(IPPs).

De nombreux efforts ont été consacrés à la découverte de ces interactions protéiques de différentes espèces tels que le système« two-hybrid »[3] et «protein co-immunoprecipitation»[4]. Les résultats sont stockés dans des bases de données publiques telles qu'IsoBase [5], NAPAbench[21], DIP[22] et IntAct[23].

La représentation de ces données sous forme de graphe aide à bien analyser les protéines où chaque protéine est représentée sous forme d'un nœud, et les interactions sous forme des arrêts entre ces nœuds [6].

Suite à l'abondance des données biologiques, la nécessité de développer de nouveaux outils et des algorithmes de calcul capable d'analyser efficacement les données de ces réseaux devenait un défi pour découvrir de nouvelles connaissances dans la recherche biologique. Une approche basée sur la comparaison de réseaux entre espèces peut constituer un cadre précieux pour accomplir ces défis.

L'alignement ou la comparaison de réseaux d'IPPs est actuellement l'un des approches les plus importantes pour aider à la compréhension du fonctionnement des organismes vivants.

Le principal objectif de l'alignement est de prédire la meilleure correspondance entre les réseaux d'IPPs. Le problème de l'alignement de réseaux d'IPPs est divisé en deux sous problèmes, l'alignement local [7, 8, 9] pour trouver les sous-réseaux conservés (similaires) localement entre les réseaux, et l'alignement global [10, 11, 12, 13, 14, 15] pour trouver la meilleure correspondance unique entre tous les nœuds de réseaux à aligner. Les deux types d'alignement peuvent être appliqués sur deux ou plusieurs réseaux d'IPPs.

Afin d'évaluer la qualité des résultats d'alignement, des métriques d'évaluation ont conçu pour vérifier la pertinence biologique de ces résultats.

Les métriques d'évaluation sont deux types : des métriques d'évaluation topologique pour vérifier la correspondance topologique des protéines alignées, et des métriques d'évaluation biologiques pour vérifier si les protéines alignées d'après l'approche d'alignement partageant les mêmes fonctions moléculaires.

La complexité informatique du problème d'alignement, est équivalente au problème d'isomorphisme de graphe qui nécessite le développement des approches heuristiques pour résoudre ce problème [19].

Bien que plusieurs méthodes heuristiques telles que LNA et GNA soient développées, il existe toujours un besoin d'améliorer la qualité de l'alignement et l'efficacité de calcul.

La plupart des approches d'alignement reposent sur les similarités séquentielles et topologiques entre les protéines pour déduire un appariement entre eux.

Les annotations GO peuvent être une mesure puissante des similarités fonctionnelles entre les protéines. Cependant, la plupart des approches d'alignement ne les utilisent que pendant l'étape de l'évaluation des qualités d'alignements et non pendant le processus d'alignement.

Dans ce cadre, nous présentons notre système d'alignement « PPINA », qui est capable d'élaborer un alignement fonctionnel en utilisant les annotations de Gene Ontology dans le processus d'alignement. Nous testons et validons les résultats de l'alignement sur de réseaux IPPs réels. La phase de validation permet de comparer les résultats obtenus par la méthode proposée avec les résultats d'autres méthodes d'alignement pour avoir une idée sur la performance de l'algorithme.

# Chapitre 1 : Les réseaux d'Interaction Protéines-Protéines

## 1.1. Introduction

Dans toutes les espèces eucaryotes, les protéines s'interagissent entre elles pour accomplir leurs fonctions moléculaires, forment un réseau d'Interaction Protéines-Protéines (IPPs).

Dans ce chapitre nous présentons quelques concepts et outils biologiques que nous exploitons tout au long de ce rapport.

## 1.2. Les protéines

Les protéines (voir Figure 1.1), sont les molécules les plus complexes et les plus variées des êtres vivants, indispensables à leur survie. Le terme protéine, du grec ancien « prôtos » signifiant premier ou essentiel, désigne une longue molécule (macromolécule) formée des acides aminés.

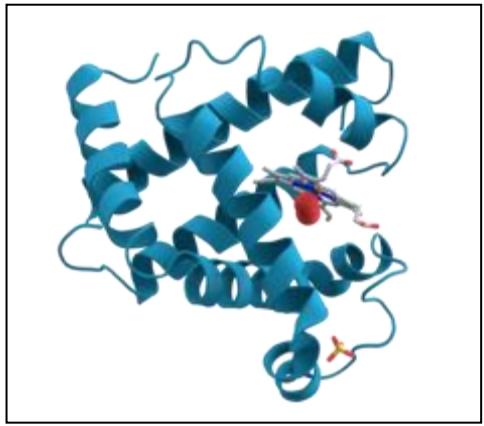

**Figure1. 1: Image 3D de la protéine myoglobin. Son rôle est transporter l'oxygène dans les tissus musculaires [31].**

L'Acide DésoxyriboNucléique(ADN), est le support stable et transmissible de l'information génétique qui définit les fonctions biologiques d'un organisme.

Lors de la méiose, l'ADN se réplique. Il est alors transcrit en Acide RiboNucléique (ARN).

L'ARN peut alors être traduit en protéines à l'aide d'une entité biologique nommée « ribosome ». On parle alors de la traduction de l'ARN en protéine (voir Figure 1.2).

On parle **des protéines homologues** si les gènes qui les codent ont une origine commune. On les reconnaît par leurs similarités présentées par leurs séquences d'acides aminés. On peut trouver des protéines homologues entre différentes espèces.

Deux **protéines sont orthologues** si elles proviennent d'espèces différentes, et **paralogues** si elles proviennent de la même espèce. On détecte informatiquement ces homologies par des algorithmes d'alignement (comme BLAST), donnant un score BLAST de similarité.

**Les complexes protéiques** sont des groupes des protéines interagissent entre eux pour accomplir une certaine fonction.

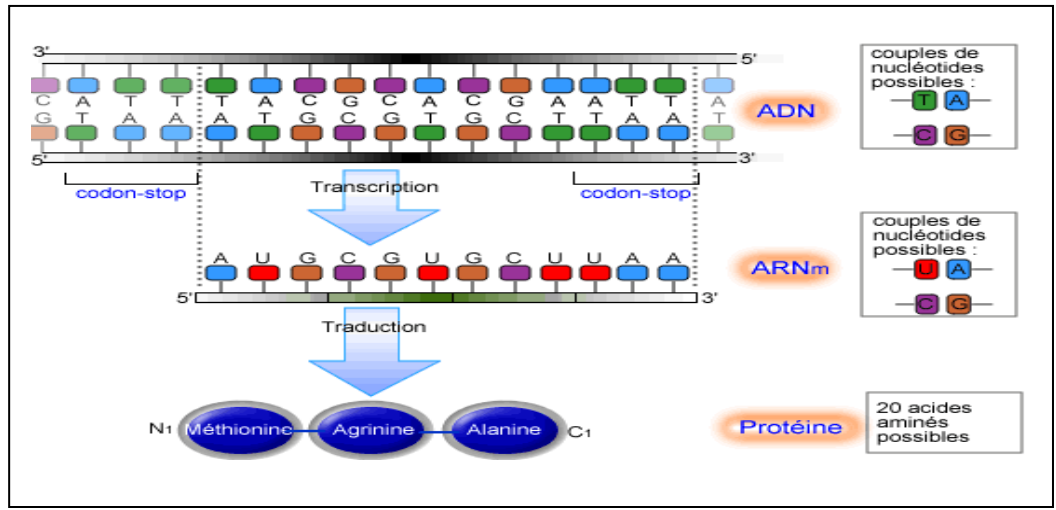

**Figure1. 2: Synthèse des protéines [32].**

## 1.3. Les interactions protéines-protéines (IPPs)

Les interactions entre les protéines sont très importantes en biologie moléculaire, et sont essentielles pour comprendre la fonction d'une protéine. Les interactions protéine-protéine(IPPs) se produisent lorsque deux protéines ou plus se lient, souvent pour accomplir leurs fonctions biologiques.

Plusieurs efforts ont été faits pour identifier ces interactions, dans le but de mieux comprendre les systèmes biologiques, tels que :

- le système two-hybrid[3],
- proteinco-immunoprecipitation[4]
- la méthode de Rosetta stone [16].

Le nombre des données sur ces interactions augmente de manière exponentielle. En 2001, les réseaux ne comportaient qu'une centaine d'interactions. De nos jours, des milliers des interactions sont disponibles grâce aux techniques listées plus haut. Cependant, toutes ces techniques souffrent d'un fort taux de bruit, il a été estimé que le nombre des **faux positifs** (interactions inscrites dans les bases de données mais n'existant pas biologiquement) est d'environ 50%. De plus, le nombre des **faux-négatifs** (interactions biologiques existantes non représentées dans les bases de données) serait de 65% [2].

Algorithmiquement, on prendra en compte ce bruit par l'ajout de poids sur les arrêts de graphe, permettant de donner un « score de confiance » à l'interaction.

### 1.4. Les réseaux d'Interaction Protéines-Protéines(IPPs)

Les réseaux d'IPPs ont été créés pour différentes espèces sur la base des interactions découvertes et publiées par des chercheurs. Les données d'un réseau d'IPPs sont représentées sous la forme d'un graphe (voir Figure1.3). Nous adoptons le terme de « réseau » lorsque l'on se rapporte aux aspects biologiques, et celui de « graphe » lorsque l'on se rapporte à l'algorithmique.

Un graphe G (V, E) où V (pour Vertex ou nœuds) est l'ensemble de ces nœuds, et E (pour Edges ou arrêtes) est l'ensemble de ses arrêts, où les nœuds symbolisent les protéines et les arrêts les interactions entre les protéines. Un graphe peut être pondéré, c'est-à-dire qu'il existe une fonction w : E→ $\mathbb{R}$ qui assigne à tout arrêt « e » de « E » un poids « w(e) », ce poids pourra donner la valeur de similarité entre deux protéines.

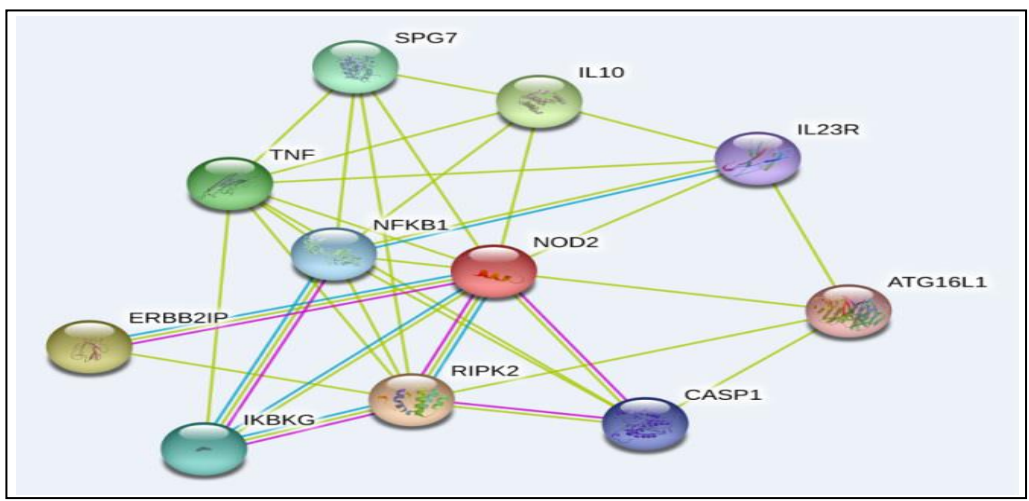

**Figure1. 3 Exemple d'un réseau d'Interaction Protéines-Protéines(IPPs)[33].**

### 1.5. Gene Ontology(Go)

Gene Ontology(Go) [17] est un vocabulaire structuré, qui décrit la fonction des gènes dans tous les organismes. Ce vocabulaire a été développé pour annoter les divers gènes disponibles et permettre la comparaison des annotations fonctionnelles des gènes conservés entre organismes.

GO est organisé en un graphe acyclique dirigé (voir Figure 1.4), où Les termes(les fonctions) sont représentés par des nœuds et les relations entre les termes sont représentées par des arrêts.

Les relations entre les termes fonctionnelles sont de type :

- « is a » : un terme A est une instance du terme parent B.
- « part of » : un terme A est une partie du terme parent B.
- « regulates » : un terme A régule le terme parent B.

Un terme peut avoir plusieurs parents et être lié à ses différents parents par des relations de type différent.

GO est divisé en trois sous-ontologies :

1. Biological Process(BP) : fait référence à l'objectif biologique auquel contribue la protéine.
2. Molecular Function (MF): fait référence à l'activité des protéines au niveau moléculaire.
3. Cellular Component (CC) : fait référence à l'emplacement des protéines dans la cellule.

GO peut être utilisé comme une mesure de similarité fonctionnelle entre les protéines, en plus il peut être utilisé comme une métrique d'évaluation pour vérifier la cohérence fonctionnelle entre les protéines alignées.

Les protéines sont fonctionnellement similaires s'ils ont des fonctions moléculaires (MF)similaires et ils sont impliqués dans des processus biologiques similaires (BP).

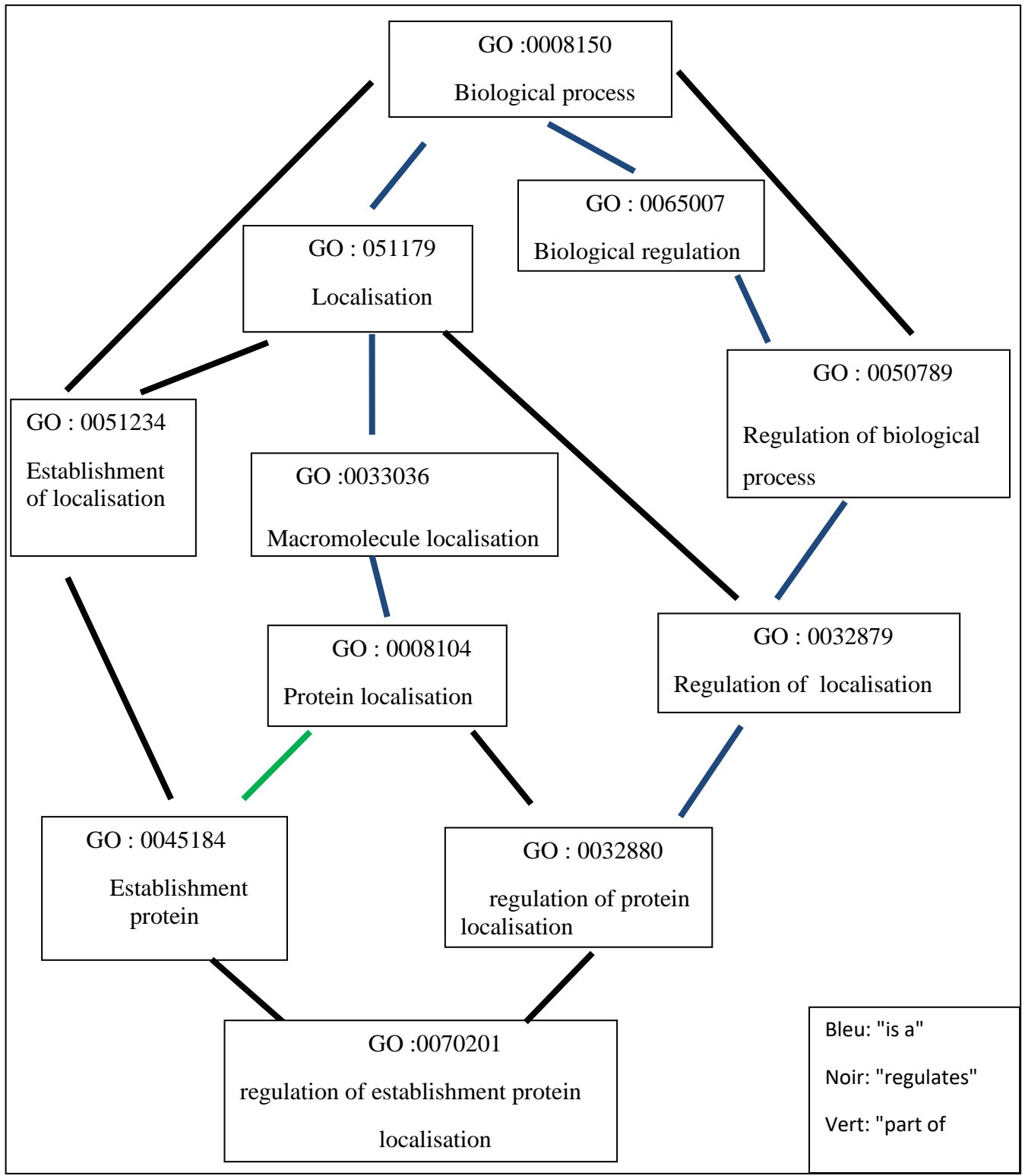

**Figure1. 4:Exemple de structure de Gene Ontology. Position dans l'ontologie "Biological Process" du terme 'regulation of establishment of protein localization'[29].**

### 1.6. Les bases de données d'IPPs

Les interactions entre les protéines découvertes par différentes méthodes tels que le système « two-hybrid »[3] et «protein co-immunoprecipitation »[4] sont stockées dans des bases de données publiques (voir tableau 1.1).

**Tableau 1. 1 un résumé de quelques bases de données des IPPs [1].**

| Sl. No. | Database | No. of species | No. of interactors | No. of interactions | Link |
| --- | --- | --- | --- | --- | --- |
| 1 | MINT | 611 | 25,530 | 125,464 | http://mint.bio.uniroma2.it/ |
| 2 | HPRD | 1 | 30,047 | 41,327 | http://www.hprd.org/ |
| 3 | BioGRID | 6 | 65,617 | 1423,105 | http://thebiogrid.org/ |
| 4 | MatrixDB | 1 | 14,533 | 26,954 | http://matrixdb.univ-lyon1.fr/ |
| 5 | IntACT | 275 | 98,289 | 720,711 | http://www.ebi.ac.uk/intact/ |
| 6 | DIP | 372 | 95,742 7 | 720,711 | http://dip.doe-mbi.ucla.edu/dip/Main.cgi |
| 7 | STRING | 2031 | 964,376 | 1380,838,440 | http://string-db.org/ |
| 8 | IsoBase | 5 | 87,737 | 114,897 | http://cb.csail.mit.edu/cb/mna/isobase/ |

### 1.7. Les métriques d'évaluation

Comme il existe un grand nombre d'alignements, et afin de déterminer le meilleur entre eux, la qualité de chaque aligneur doit être évaluée.

### 1.7.1. Les métriques d'évaluation topologique

#### 1.7.1.1. EC (Edges Correctness)

« EC » renvoie le pourcentage des arrêts corrects du premier réseau qui sont alignés aux arrêts de l'autre réseau. Plus la valeur de « EC » est élevée cela indique que les deux réseaux sont topologiquement similaires [12].

$$EC = \frac{|\{(u,v) \in E_1 : g(u), g(v) \in E_2\}|}{|E|} X\ 100\% \tag{1}$$

Où (u, v) est l'interaction entre deux protéines de premier réseau, g(u), g(v) leur similaire dans le deuxième réseau et « E » désigne le minimum nombre d'interactions entre les deux réseaux.

### 1.7.2. Les métriques d'évaluation biologique

Pour mesurer l'uniformité fonctionnelle de chaque paire des nœuds alignés, nous avons testé la cohérence des annotations GO pour vérifier si les protéines prédites ont les mêmes fonctions moléculaires.

Dans ce cadre, nous avons appliqué les métriques Mean Entropy(**ME**) et le Mean Normalized Entropy(**MNE**) qui comparent les termes fonctionnels de GO entre les protéines pour vérifier leurs cohérences fonctionnelles. Les valeurs les plus inferieures d'entropie indiquent une cohérence fonctionnelle plus élevée entre les protéines [29].

#### 1.7.2.1. Mean Entropy (ME)

$$H(S_v^*) = H(p_1, p_2, \ldots, p_d) = -\sum_{i=1}^{d} p_i log p_i \tag{2}$$

Où $p_i$ est le pourcentage de toutes les protéines ont des termes i de GO et d est le nombre total des termes GO.

### 1.7.2.2. Mean Normalized Entropy (MNE)

$$\bar{H}(S_v^*) = \frac{1}{\log d} H(S_v^*) \tag{3}$$

## 1.8. Conclusion

Nous avons présenté dans ce chapitre, des notions et des outils biologiques, nécessaires à la compréhension et à l'évaluation de processus d'alignement de réseaux d'IPPs.

Voici un résumé des principaux points que nous avons abordés :

-L'analyse des interactions entre les protéines est essentielle pour comprendre leurs fonctions.

-Les réseaux d'IPPs sont représentés sous forme d'un graphe G=(V, E), où V symbolise les nœuds(protéines) et E symbolise les arrêts(interactions) entre les nœuds.

-Les résultats des approches d'alignements doivent être évalués pour vérifier la cohérence biologique entre les nœuds alignés.

# Chapitre 2 : État de l'art sur l'alignement de réseaux d'Interaction Protéines-Protéines(IPPs)

## 2.1 Introduction

L'alignement de réseaux d'IPPs est actuellement l'un des approches les plus importantes pour comprendre le fonctionnement des organismes vivants.

Le processus d'alignement vise à répondre à des questions biologiques :

Quelles protéines ? Et quels groupes d'interactions protéiques de différentes espèces sont susceptibles d'avoir des fonctions similaires ?

Dans ce chapitre, nous étudions l'alignement de réseaux d'IPPs, nous commençons tout d'abord par définir les divers types de l'alignement et enfin nous nous intéressons à quelques approches existantes.

## 2.2 L'alignement de réseaux d'IPPs

L'alignement ou la comparaison de réseaux d'IPPs de différentes espèces vise à trouver une meilleure correspondance entre les nœuds de différents réseaux d'IPPs. Formellement, étant donné deux réseaux d'entrée G1 et G2, le problème d'alignement consiste à rechercher des correspondances entre les nœuds de G1 et les nœuds de G2.

L'alignement de réseaux d'IPPs se divise en deux types :

- L'alignement d'une paire de réseaux d'IPPs.
- L'alignement de multiples réseaux d'IPPs.

### 2.2.1 L'alignement par paire

La plupart des travaux d'alignement de réseaux IPPs se concentrent sur l'alignement par paire [7, 8, 9, 12, 13, 15]. L'alignement par paire est limité à deux réseaux d'IPPs, l'objectif de comparer deux réseaux de différentes espèces est de trouver les protéines homologues.

### 2.2.2 L'alignement multiple

Bien que plusieurs approches d'alignement de réseaux soient conçues pour gérer plus que deux réseaux d'entrées, ils peuvent être utilisés comme des aligneurs par paire lorsque le nombre de réseaux d'IPPs n'est que deux réseaux.

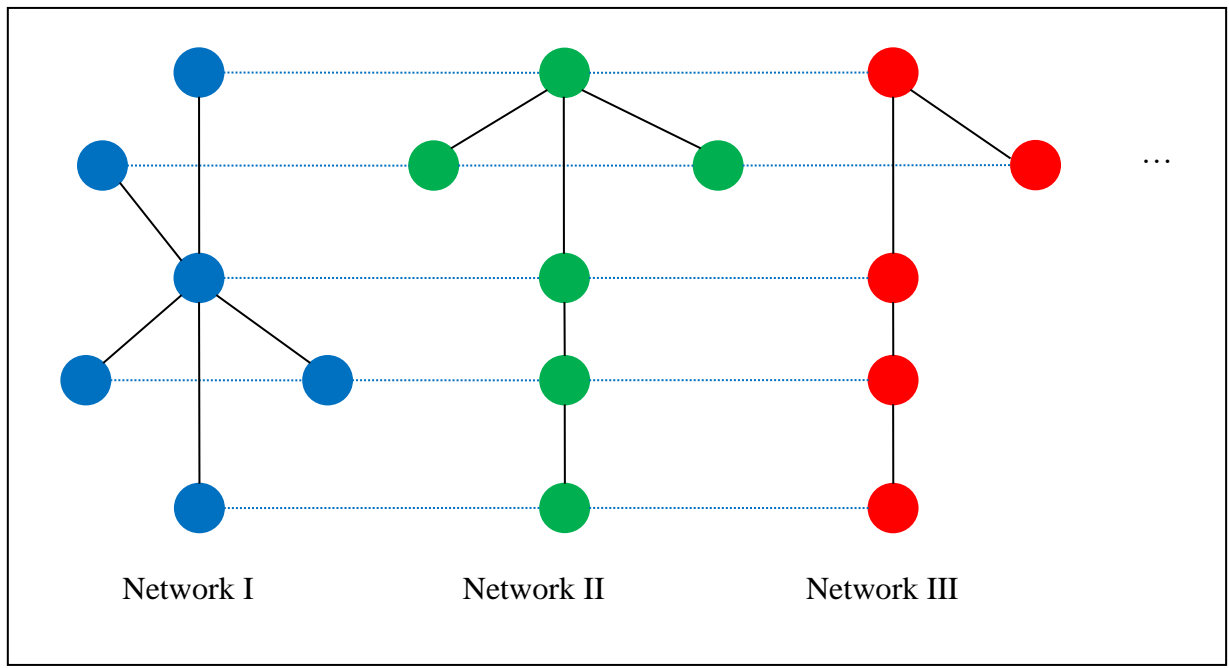

**Figure 2. 1: Exemple d'alignement multiple de trois réseaux.**

Différentes méthodes d'alignement par pairs et multiples sont disponibles pour les réseaux IPPs, elles peuvent se classifier en des méthodes d'alignement locales et globales.

### 2.2.3 L'alignement global et local

Les deux types d'alignement de réseaux peuvent aligner deux ou multiples réseaux. L'alignement local cherche localement les sous-réseaux les plus similaires entre les réseaux à aligner. Tandis que l'alignement global cherche une seule grande région avec une correspondance individuelle entre toutes les protéines de différentes espèces.

L'alignement global produit un alignement one-to-one, c'est-à-dire que chaque nœud aune seule correspondance tandis que l'alignement local produit un alignement many-to-many, c'est-à-dire qu'il trouve les régions les plus similaires entre les deux réseaux comparés (voir Figure 2.2).

Il y a un lien entre l'alignement local et global : les deux visent à trouver des similitudes topologiques et fonctionnelles entre les réseaux comparés afin de permettre le transfert des connaissances biologiques entre les espèces.

Par conséquent, la question est de savoir lequel utiliser: une approche d'alignement local ou global ou une approche qui réconcilierait les deux?

Les analyses [19] ont montré que la qualité de l'alignement local est fonctionnellement mieux que topologiquement tandis que la qualité de l'alignement global est topologiquement mieux que fonctionnellement. Autrement dit, l'alignement local a surpassé l'alignement global fonctionnellement et l'alignement global a surpassé l'alignement local topologiquement. Dans ce cadre, une approche a été proposée comme IGLOO [20] qui vise à hériter la haute qualité fonctionnelle de l'alignement local et la haute qualité topologique de l'alignement global.

Par conséquent, la réconciliation entre ces deux types d'alignement est un problème de recherche ouverte qui devrait être étudiée de manière approfondie à l'avenir.

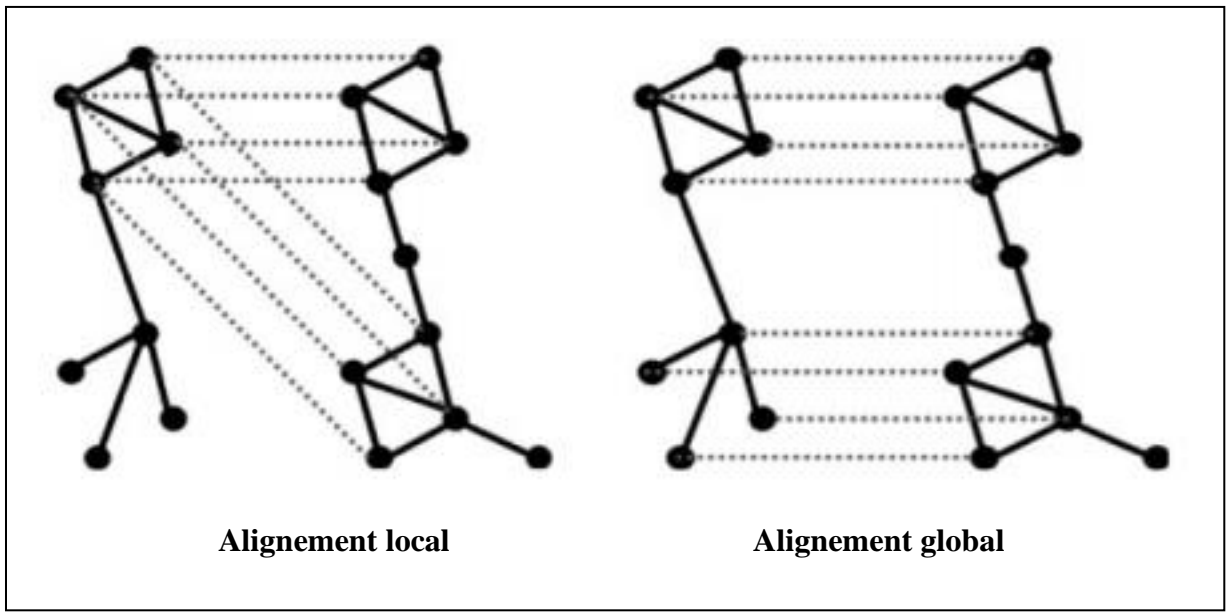

**Alignement local**      **Alignement global**

**Figure 2. 2 : Alignement local versus alignement global de deux réseaux. Les nœuds alignés les uns aux autres entre les 2 réseaux à comparer sont indiqués par des pointillés [19].**

Dans cette section, nous discutons différentes méthodes d'alignement global et local entre deux ou plusieurs réseaux d'IPPs de différentes espèces.

## 2.3 Les approches d'alignements locaux de réseaux d'IPPs

### 2.3.1. PathBlast

PathBlast [7] est un aligneur local d'une paire de réseaux d'IPPs. Comme BLAST produit un alignement local de séquences des protéines, PathBlast produit un alignement local de réseaux d'IPPs.

L'algorithme PathBlast fonctionne sur deux réseaux d'IPPs pour identifier les chemins d'interactions protéiques conservées (similaires). PathBlast fait référence au premier réseau en tant que requête et au deuxième réseau en tant que cible, PathBlast génère tous les chemins de la cible ayant des scores d'alignement élevés avec la requête.

Pour définir la requête de chemin, les utilisateurs doivent entrer une série d'Ids des protéines ou une série des séquences protéiques, et les aligner contre un réseau d'IPPs sélectionné de la base des données de réseaux où le nombre de chemin entré peut varier de deux à cinq protéines.

La méthode cherche des alignements avec des scores élevés entre les pairs des chemins d'interactions protéiques. Les scores d'alignements des chemins sont calculés en fonction des similarités séquentielles et la qualité des interactions protéiques.

### 2.3.2. NetworkBLAST

NetworkBLAST [8] est un aligneur local d'une paire de réseaux d'IPPs qui identifie les complexes protéiques conservés contrairement à PathBlast qui détecte des chemins conservés (similaires).

NetworkBLAST traite deux réseaux d'IPPs en intégrant les similarités séquentielles des protéines. Le score de chaque complexe protéique est calculé en fonction d'un modèle de complexe protéique « a log likelihood ratio score », et les complexes en sortie sont triés en fonction de leurs scores.

Les complexes identifiés peuvent être utilisés pour prédire la fonction des protéines et leurs interactions.

### 2.3.3. NetAligner

NetAligner (Network Aligner) [9] est un aligneur local d'une paire de réseaux d'IPPs pour identifier les complexes protéiques conservés et prédire les interactions conservées. À partir de deux réseaux d'entrée NetAligner commence par créer le graphe initial composé de protéines homologues collectées de deux réseaux et les interactions issues des «interactomes» des espèces.

Les « Seeds » sont identifiés à partir des composants détectés de l'alignement de ce graphe, puis seront étendus à d'autres « Seeds » en utilisant « gaps » ou « mismatch edges » et les composants du graphe étendu sont l'alignement final.

NetAligner est un serveur web où l'utilisateur peut sélectionner l'une des trois tâches « Complex to interactome », « Pathway to interactome » ou « Interactome to interactome ».

Dans le cas du « Complex to interactome », la requête est un complexe peut-être aligné à « l'interactome » du même ou de différentes espèces.

Dans la tâche « Pathway to interactome », la requête est un chemin peut-être aligné à « l'interactome » du même ou de différente espèce.

La tâche « Interactome to interactome », aligne tout l'ensemble des « interactomes » de deux espèces pour trouver les complexes protéiques ou les sous-réseaux.

## 2.4. Les approches d'alignements globaux de réseaux d'IPPs

### 2.4.1. IsoRankN

IsoRankN ou IsoRank-Nibble [11] est une évolution de l'algorithme originale « IsoRank ». Il résout le problème de l'alignement global de multiples réseaux et peut être utilisé comme un aligneur par paire lorsque seulement deux réseaux sont donnés en entrée.

Le fonctionnement d'IsoRankN commence par calculer les scores des similarités fonctionnelles de chaque paire des protéines de deux espèces différentes en utilisant algorithme « IsoRank » originale sans effectuer leur étape finale qui est générée en utilisant l'algorithme de partitionnement spectral d'IsoRankN.

### 2.4.2. SMETANA

SMETANA (Semi-Markov randomwalk scores Enhanced by consistency Transformation for Accurate Network Alignment)[10] est un aligneur global de multiples réseaux pour trouver l'alignement maximum attendu (Maximum Expected Accuracy (MEA)).

Dans la première étape, à l'aide d'un modèle de marche aléatoire semi-markovienne les scores de similarité qui indiquent la probabilité d'alignement de deux nœuds de réseaux différents sont estimés. Après, deux transformations de cohérence probabilistes sont utilisées pour améliorer les probabilités d'alignement de toute paire de nœuds calculés à la première étape :

La première transformation est la transformation de cohérence probabiliste intra-réseau qui intègre des informations sur les voisins du nœud, pour mettre à jour les scores calculés dans la première étape. La seconde transformation est connue sous le nom de cohérence probabiliste inter-réseaux qui intègre des informations provenant d'autres réseaux dans l'alignement.

Dans la deuxième étape, les scores transformés sont ensuite utilisés pour construire l'alignement global MEA.

### 2.4.3. NETAL

NETAL (NETwork Alignement) [12] est un aligneur global d'une paire de réseau d'IPPs. L'approche fait l'alignement de deux réseaux d'IPPs en passant par deux phases, la première phase consiste à construire la matrice des scores d'alignement et l'utiliser dans la deuxième phase pour trouver l'alignement global final.

La matrice des scores d'alignement est construite sur la base de deux autres matrices, la matrice des scores des similarités et la matrice des scores d'interactions.

La matrice des scores des similarités est construite en se basant sur la matrice des scores topologiques qui indique la similarité topologique entre les nœuds de deux réseaux à aligner.

Les scores des similarités topologiques sont calculés sous l'hypothèse que deux nœuds sont topologiquement similaires si et seulement si leurs voisins sont topologiquement similaires.

La matrice des scores d'interactions indique une valeur approximative des nombres d'interactions conservées. Ainsi, en combinant la matrice des similarités et la matrice d'interactions, une matrice des scores d'alignement est construite pour déduire l'alignement global final.

L'algorithme commence par aligner les nœuds avec le meilleur score d'alignement, puis mettre à jour la matrice d'interaction, et par conséquent, mettre à jour la matrice des scores d'alignement. Ce processus se répète jusqu'à ce que tous les nœuds de réseaux ayant le minimum nombre des nœuds soient alignés sur les nœuds de l'autre réseau.

### 2.4.4. PINALOG

PINALOG [13] est un aligneur global d'une paire de réseaux d'IPPs qui combine les informations biologiques et topologiques.

PINALOG aligne deux réseaux d'IPPs en trois étapes. L'idée de la première étape « community detection » est que les communautés dans les réseaux d'IPPs indiquent souvent des groupements fonctionnels des protéines dans les réseaux. Pour identifier ces communautés PINALOG utilisait la méthode « CFinder ».

La deuxième étape « community mapping » est déterminée en deux sous étapes : la première sous étape consiste à créer une matrice des scores des similarités de toutes les communautés extraits, et l'utiliser dans la deuxième sous étape pour obtenir la meilleure correspondance entre ces communautés. Les scores de similarité sont une combinaison entre les scores séquentiels et fonctionnels des protéines.

Il faut noter que la similarité séquentielle de deux protéines est calculée sur la base de leurs scores BLAST où BLAST (ui, vj) est le score BLAST entre les deux protéines ui et vj. En plus, la similarité fonctionnelle de deux protéines annotée avec des termes GO est calculée par la méthode Schlicker [18].

La troisième étape « extension mapping » est réalisée en considérant les voisins des protéines comme candidats pour étendre l'alignement en intégrant les similarités topologiques sous la forme de « neighbourhood similarity ».

### 2.4.5. MAPPIN

MAPPIN (Multiple Alignment of Protein Protein Interaction Network) [15] est une autre approche d'alignement multiple global de réseaux d'IPPs et peut fonctionner à la fois comme un aligneur par paires si le nombre des réseaux ne dépasse pas deux réseaux d'IPPs. MAPPIN utilise les informations topologiques, séquentielles, et fonctionnelles pour aligner les réseaux d'IPPs.

L'algorithme commence par donner à chaque arrêt du graphe biparti un poids calculé en utilisant les informations de GOA(Gene Ontology Annotation) et les informations séquentielles pour chaque protéine alignée. L'étape suivante consiste à collecter les « seed » avec les meilleurs scores à partir de graphe bipartiaprès, chaque « seed » est développé de manière itérative en explorant « the local neighborhood » pour chaque protéine comparée.

À l'étape finale MAPPIN applique la fonction « Simulated Annealing(SA) » afin de trouver l'alignement global.

### 2.4.6. MI-GRAAL

MI-GRAAL (Matching-based Integrative GRAph Aligner) [24] est un aligneur d'une paire de réseaux d'IPPs qui intègrent des métriques des similarités topologiques et biologiques pour aligner les nœuds de deux réseaux.

MI-GRALL intègre cinq métriques de similarité : quatre mesures de similarités topologiques et une mesure de similarités biologiques.

Les mesures des similarités topologiques sont : (1) « Graphletdegree signature distance (SD) », (2) « Relative degreedifference (DD) », (3) «Relative clustering coefficient difference (CD) », (4) « Relative eccentricitydifference (ED) ». La cinquième métrique est les scores des similarités séquentielles mesurées par BLAST.

MI-GRAAL construit cinq matrices des similarités indépendantes en se basant sur ces cinq métriques des similarités et les combinant en une seule matrice de confiance « C ». En se basant sur les scores de la matrice « C » MI-GRAAL construit une file d'attente des nœuds dans l'ordre décroissant. Après avoir aligné les nœuds MI-GRAAL utilise «the maximum weight bipartite matching » pour aligner ensuite les voisins des nœuds alignés. Enfin MI-GRAAL utilise la méthodologie « seed-and-extend » pour trouver l'alignement final.

### 2.4.7. MAGNA++

MAGNA++(Maximizing Accuracy in Global Network Alignment)[30], est un aligneur global d'une paire de réseaux d'IPPs. MAGNA++est une amélioration de la version originale de l'algorithme MAGNA, qui maximise la conservation des nœuds et des arrêts pour améliorer la qualité de l'alignement par combiner entre la mesure de conservation des nœuds et la mesure de conservation des arrêts.

MAGNA++ utilise la similarité de « graphlet degree vector » comme une mesure de conservation des nœuds et peut utiliser d'autres mesures telles que la similarité séquentielle entre les protéines. De plus, afin de réduire le temps de l'exécution MAGNA++ utilise le « multi-threading » et implémente les mesures de conservation des arrêts de l'algorithme original MAGNA.

## 2.5. Une approche locale et globale

### 2.5.1. IGLOO

IGLOO (Integrating global and local biological network alignement)[20] est une méthode développée pour améliorer la qualité de l'alignement par combiner entre les principes d'alignement local et global. Le processus d'alignement d'IGLOO comprend quatre étapes.

IGLOO commence par générer un alignement local initial de deux réseaux d'IPPs par une approche d'alignement local existante.

Dans la deuxième étape IGLOO construit une fonction pour calculer les scores des similarités entre les nœuds de deux réseaux appelés « node cost function » en utilisant la méthode de NETAL [12], cette fonction sera utilisée dans la troisième et la quatrième étape pour modifier l'alignement initial généré dans la première étape.

La troisième étape du processus d'alignement d'IGLOO consiste à supprimer les nœuds de l'alignement initial qui ont les scores des similarités les plus bas. Le processus de suppression s'arrête lorsque les paires des nœuds alignés restants dans l'alignement sont égaux à la taille de l'alignement précis (la taille de l'alignement est donnée par l'utilisateur).

La dernière étape d'IGLOO consiste à ajouter les paires des nœuds non alignés qui ont les scores de similarités les plus élevés jusqu'à ce qu'aucune paire des nœuds ne peut pas être ajoutée et après cette étape, l'alignement trouvé est considéré comme l'alignement final.

Le tableau (2.1) résume les approches d'alignement de réseaux d'IPPs discutées dans ce rapport.

**Tableau 2. 1: un résumé de différentes approches d'alignement de réseaux IPPs**

| Alignement | Aligneurs | année | Alignement par paire | Alignement multiple | Approche topologique | Approche fonctionnelle | Méthodologies |
|---|---|---|---|---|---|---|---|
| Local | PathBlast | 2004 | ✔ | ✘ | ✔ | ✔ | les similarités BLAST des protéines et les probabilités d'interactions |
| | NetworkBLAST | 2008 | ✔ | ✘ | ✔ | ✔ | a log likelihood ratio score |
| | NetAligner | 2012 | ✔ | ✘ | ✔ | ✘ | 'gaps' ou ' mismatch edges' |
| Global | IsorankN | 2009 | ✘ | ✔ | ✔ | ✔ | La similarité des voisins. |
| | PINALOG | 2012 | ✔ | ✘ | ✔ | ✔ | La méthode CFinder |
| | SMETANA | 2013 | ✘ | ✔ | ✔ | ✔ | Semi-Markov random walk scores. |
| | NETAL | 2013 | ✔ | ✘ | ✔ | ✘ | La similarité des voisins. |
| | MAPPIN | 2018 | ✔ | ✔ | ✔ | ✔ | Simulated Annealing(SA) |
| | MAGNA++ | 2015 | ✔ | ✘ | ✔ | ✔ | Maximiser la conservation des nœuds et des arrêts. |
| | MI-GRAAL | 2011 | ✔ | ✘ | ✔ | ✔ | Intégrer plusieurs métriques de similarité topologique. |
| Local et Global | IGLOO | 2016 | ✔ | ✘ | ✔ | ✔ | Intégrer plusieurs métriques de similarité topologique. |

## 2.6. Les domaines d'applications de l'alignement de réseaux d'IPPs

Les algorithmes d'alignements de réseaux sont l'un des outils les plus puissants pour comparer les réseaux IPPs. L'alignement de réseaux IPPs a de nombreuses applications tels que :

- La prédiction des fonctions des protéines [24].
- La reconstruction des arbres phylogénétiques [16, 24].
- La prédiction des interactions entre les protéines [16].
- La détection des orthologues [25].
- La conception des médicaments [26].
- Le transfert des connaissances sur le fonctionnement cellulaire d'espèces étudiées vers des espèces non étudiées [6].
- Améliorer la compréhension des maladies infectieuses [28].

## 2.7. Conclusion

Dans ce chapitre, nous avons défini l'alignement de réseaux d'IPPs tout en détaillant les différences entre les différents types d'alignements. Enfin, nous avons essayé de présenter brièvement quelques approches existantes d'alignement de réseaux d'IPPs.

# Chapitre 3 : La Conception d'un algorithme pour l'alignement global de réseaux d'IPPs

## 3.1. Introduction

Dans la biologie informatique, l'alignement de réseaux d'IPPs représente un grand intérêt pour gérer les informations biologiques. Les chercheurs du domaine découvraient plusieurs méthodes d'alignement de réseaux d'IPPs. Ces méthodes se basent sur la similarité entre les protéines de réseaux de différentes espèces, et souvent exploitées dans le domaine biologique pour découvrir de nouvelles connaissances biologiques.

Dans ce chapitre, nous allons décrire l'algorithme« PPINA » pour l'alignement global d'une paire de réseaux d'IPPs. L'algorithme aligne deux réseaux d'IPPs de différentes espèces en se basant sur la similarité topologique, séquentielle et fonctionnelle des protéines pour déduire un alignement global entre eux.

Nous commençons par définir quelques notions de l'alignement, et ensuite nous allons détailler les différentes étapes de notre alignement.

## 3.2. Notions de l'alignement global de deux réseaux d'IPPs

Etant donné deux réseaux d'IPPs $G1= (V1, E1)$ et $G2= (V2, E2)$ où $|V1|$, $|V2|$ représentent le nombre des nœuds (protéines) et $|E1|$, $|E2|$ représentent le nombre des arrêts (interactions) non-orienté entre les nœuds respectivement pour G1 et G2.

Soit $i \in V1$ et $j \in V2$, $e= (i, j)$ est l'interaction entre le nœud « i » et « j » où l'interaction est pondérée par un score de similarité des deux nœuds et $0 \leq w(e) \leq 1$.

N(i) est l'ensemble des voisins de nœud« i »(un voisin est le nœud qui est en interaction directe avec un autre nœud dans le même réseau).

Soit $i, u \in V1$ et $j, v \in V2$, où u est le voisin de « i » et « v » est le voisin de « j », si « i » est aligné avec « j » et « u » est aligné avec « v » alors l'interaction (i, u) est alignée avec l'interaction (j, v) donc l'interaction (i, u) est dite une interaction conservée (similaire)(voir Figure 3.1).

On suppose que $|V1| \leq |V2|$, l'alignement global de deux réseaux est une fonction injective f : V1➔ V2, vise à trouver le meilleur alignement entre tous les alignements possibles de chaque nœud de |V1| à |V2|.

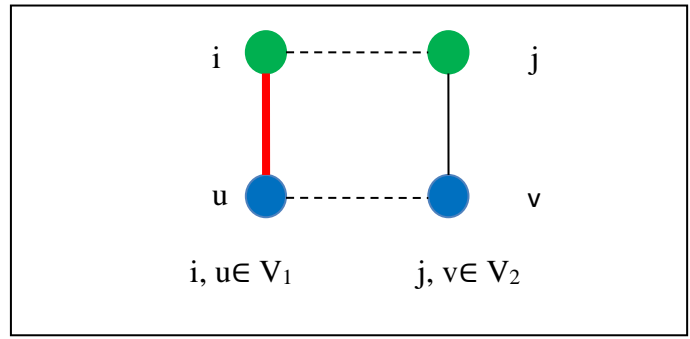

**Figure3. 1: Exemple d'une interaction conservée**

Nous proposons une approche qui permet d'aligner une paire de réseaux d'IPPs en combinant entre les informations topologiques, séquentielles et fonctionnelles entre les protéines.

Cette approche est une extension de l'algorithme « NETAL » [12] décrit dans le chapitre 2.

« NETAL » aligne deux réseaux d'IPPs topologiquement seulement. Tandis que « PPINA »aligne deux réseaux d'IPPs topologiquement, séquentiellement et fonctionnellement.

## 3.3. Le processus d'alignement de réseaux d'IPPs de« PPINA »

« PPINA » prend comme entrée deux réseaux d'IPPs G1 ($V_1$, $E_1$), G2 ($V_2$, $E_2$), les « BLAST scores » pour calculer les scores des similarités séquentielles entre les protéines, les annotations des gênes pour calculer les scores des similarités fonctionnelles et des paramètres de configuration et prédit un alignement global entre G1 et G2.

Le processus d'alignement présenté par la (Figure 3.2), passe par deux étapes afin de prédire une correspondance entre les protéines de deux réseaux :

1. La construction de la matrice des scores d'alignement.
2. L'alignement global final de deux réseaux.

Nous découvrons ces deux étapes en détail par la suite.

« PPINA » calcule la similarité entre chaque protéine i $\in V_1$ et j $\in V_2$ en utilisant les informations topologiques, séquentielles et fonctionnelles des protéines.

La matrice des scores des similarités est calculée en se basant sur ces trois informations biologiques.

Le score d'interaction entre deux protéines i ∈ $V_1$ et j ∈ $V_2$ est une estimation de nombres d'interactions conservées incidente de « i » en supposant que « i » et « j » sont alignés.

La matrice des scores des similarités calculée durant la première étape reste inchangée jusqu'à la fin du processus d'alignement. La raison est que la matrice des similarités est calculée sur la base de la structure topologique de réseaux et les propriétés biologiques qui sont disponibles au début du processus.

Cependant, la matrice des scores d'interactions doit être mise à jour de manière itérative pendant la deuxième phase de l'algorithme car une fois deux protéines sont alignées, ils seront supprimés de la matrice des scores d'alignement, et par conséquent, le nombre d'interactions conservées des autres protéines change dans la matrice des scores d'interactions. Ainsi la matrice d'alignement doit être mise à jour respectivement.

La deuxième étape consiste à chercher les protéines non alignés ayant le maximum de scores d'alignement et les aligner. Cette étape est répétée jusqu'à ce que toutes les protéines du premier réseau soient alignées sur les protéines du deuxième réseau.

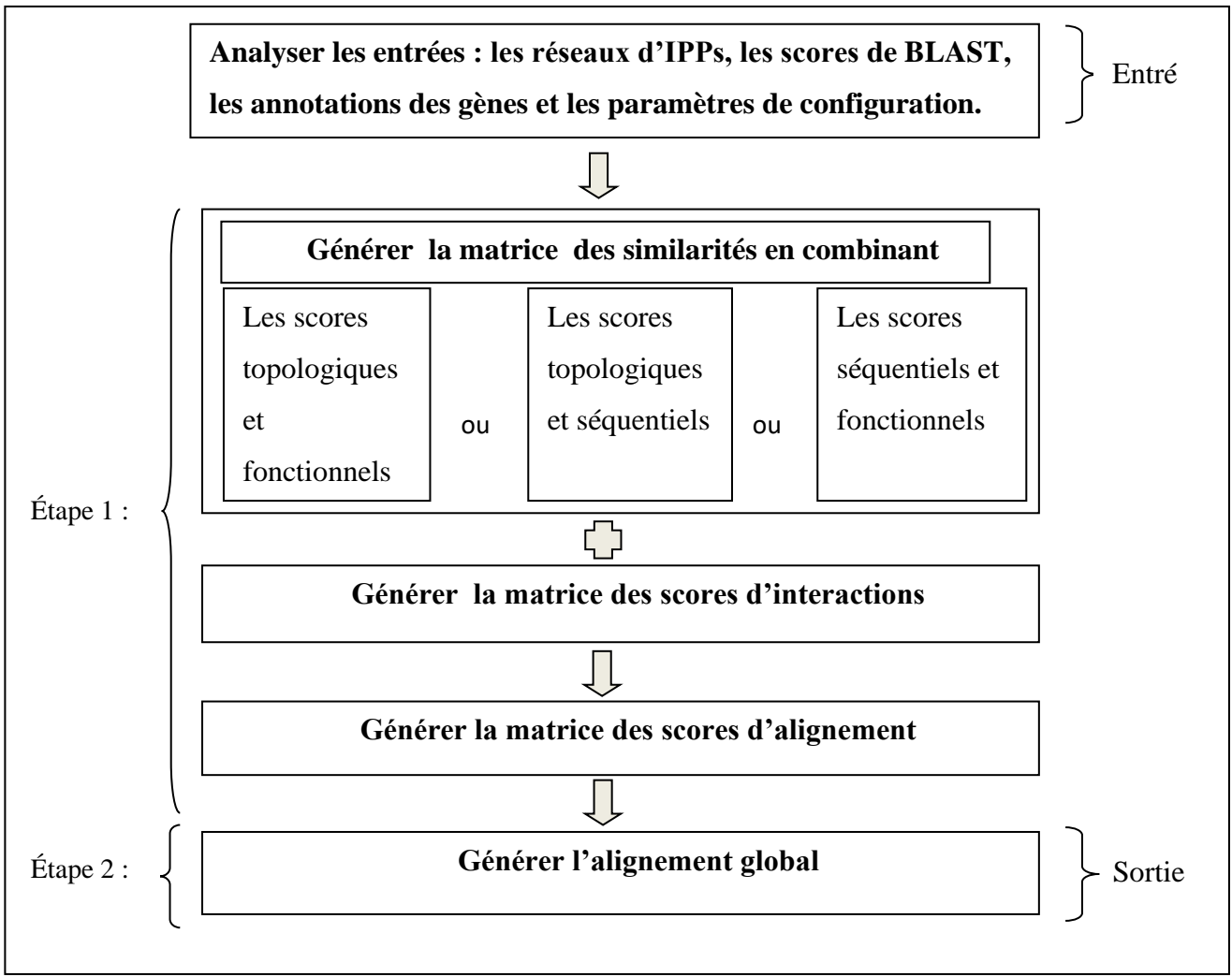

**Figure3. 2: Le processus d'alignement global de «PPINA ».**

Le processus d'alignement de réseaux d'IPPs est composé de deux étapes :

1. La construction de la matrice des scores d'alignement.

2. L'alignement global final de deux réseaux.

Dans cette section, nous allons détailler ces deux étapes.

### 3.3.1. La première étape : la construction de la matrice des scores d'alignement

La matrice des scores d'alignement est calculée sur la base de deux autres matrices :

- la matrice des scores des similarités
- la matrice des scores d'interactions.

#### 3.3.1.1. La matrice des scores des similarités

La matrice des scores des similaités « S » est constituée de |V₁| lignes et |V₂| colonnes, S(i,j) indique le score de similarité entre les protéines i∈V₁ et j∈V₂ des deux réseaux G1 et G2.

Dans ce travail, nous calculons la matrice des scores des similarités par la combinaison des deux types des similarités, soit :

- Une combinaison entre la matrice des scores des similarités topologiques et fonctionnels des protéines :

$$S(i,j) = \alpha T(i,j) + (1-\alpha) * F(i,j)$$

- Une combinaison entre la similarité des scores des similarités topologiques et séquentiels des protéines :

$$S(i,j) = \alpha T(i,j) + (1-\alpha) * B(i,j)$$

- Une combinaison entre la similarité des scores des similarités fonctionnels et séquentiels des protéines :

$$S(i,j) = \alpha B(i,j) + (1-\alpha) * F(i,j)$$

Où T(i, j) présente la valeur de la matrice des similarités topologique, F(i, j) présente la valeur de la matrice de similarité fonctionnelle et B(i, j) présente la valeur de la matrice de similarité séquentielle respectivement pour i∈V₁ et j∈V₂ et 0≤α≤1, un paramétre qui équilibre entre les valeurs des deux matrices. Les valeurs des matrices topologiques, fonctionelles et biologiques reste inchangé jusqu'à la fin du processus .

#### 3.3.1.1.1. La matrice des scores des similarités topologiques

La matrice des scores des similarités topologique « T » est constitué de |V₁| lignes et |V₂| colomnes, T(i,j) indique la similarité topologique entre les protéines i ∈V₁ et j∈ V₂. La similarité topologique entre deux protéines est calculée en se basant sur les voisins de chaque protéine où deux protéines sont topologiquement similaires si et seulement si leurs voisins sont topologiquemnt similaires.

On note N(i) et N(j) le nombre des voisins respectivement pour i ∈ $V_1$ et j ∈ $V_2$.

Pour calculer le score topologique de toutes les protéines de G1 et G2, nous initialisons toutes les valeurs de la matrice à1 $T^0(i, j)= 1$ et dans chaque itération $T^{t+1}(i, j)$ on construit un graphe biparti $G_b= (V_b, E_b)$ où $V_b$ présente deux ensembles disjoints des voisins des deux protéines à comparer N(i) et N(j) et $E_b$ présente l'ensemble des interactions entre ces voisins (i', j')∈$E_b$ où i'∈ N(i) et j'∈ N(j).

Après la construction de $G_b$, on cherche le nombre des voisins communs entre les deux protéines à aligner, pour ce faire on selecte un arrêt e= (u, v), on ajoute « e » à l'ensemble M, on supprime « u » et « v » et tous les arrêts incidentent de « u » et « v » dans $G_b$ et on répète ce processus jusqu'à ce que tous les arrêts entre les nœuds seront supprimés. Après, on calcule $T_{t+1}(i, j)$ en utilisant la formule(4) :

$$T^{t+1}(i,j) = \frac{\sum_{(u,v)\in M} T^t(u,v)}{\max\{|N(i)|, |N(j)|\}} \qquad (4)$$

Où « t » est le compteur d'iteration, l'enumérateur est la somme des scores topologique des interactions de l'ensemble M, qui présente le minimum nombre des voisins min{|N(i)|, |N(j)|} et le dénominateur présente le maximum nombre des voisins respectivement entre lesprotéines i ∈ $V_1$ et j ∈ $V_2$.

### 3.3.1.1.1.1. Exemple de calculer le score topologique entre deux protéines de deux réseaux G1 et G2

Etant donné deux réseaux d'IPPs G1 et G2 et la matrice de leurs scores topologique à l'état initiale T0 :

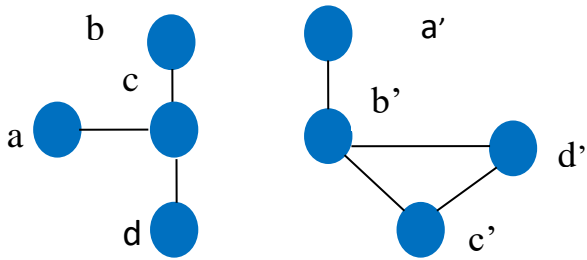

|   | a' | b' | c' | d' |
|---|----|----|----|----|
| **a** | 1 | 1 | 1 | 1 |
| **b** | 1 | 1 | 1 | 1 |
| **c** | 1 | 1 | 1 | 1 |
| **d** | 1 | 1 | 1 | 1 |

**T0**

Supposons qu'on veuille calculer le score topologique de T (c, b') et T(c, c') :

- N(c)= {a, b, d}, N(b')= {a', c', d'}, N (c')= {b', d'}
- Le maximum nombre de voisins entre les deux nœuds c et b'= 3

- Le maximum nombre de voisins entre les deux nœuds c et c'= 3

**T (c, b')= ?**

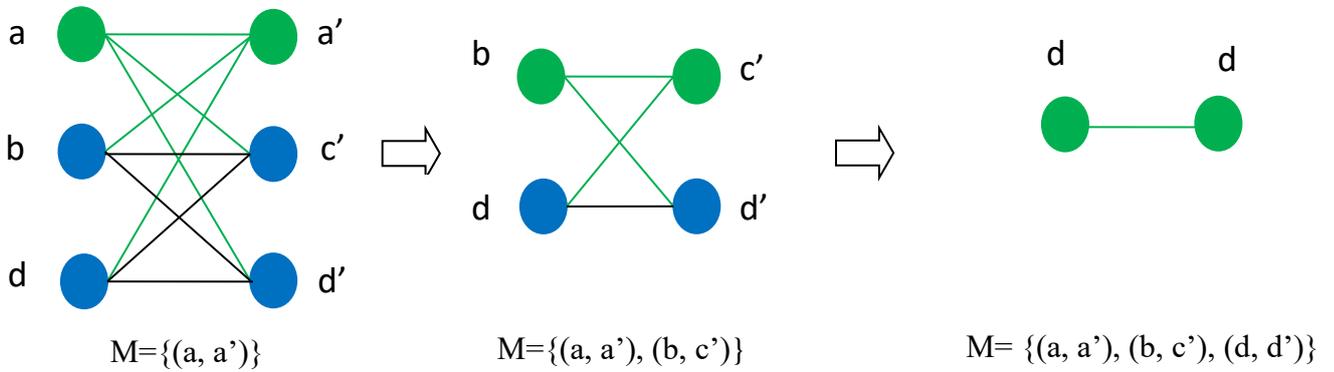

M={(a, a')}    M={(a, a'), (b, c')}    M= {(a, a'), (b, c'), (d, d')}

$$T^2 (c, b') = \frac{T(a,a') + T(b,c') + T(d,d')}{3} = \frac{3}{3} = 1$$

**T (c, c')= ?**

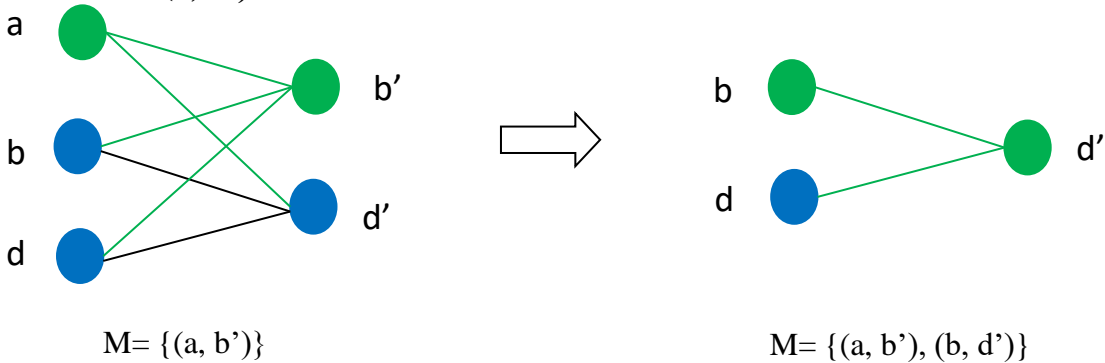

M= {(a, b')}    M= {(a, b'), (b, d')}

$$T^3 (c, c') = \frac{T(a,b') + T(b,d')}{3} = \frac{2}{3} = 0.66$$

Après t itérations la matrice finale des scores des similarités topologique entre toutes les protéines de G1 et G2 est dans le tableau ci-dessous :

**T**

|   | a'   | b'   | c'   | d'   |
|---|------|------|------|------|
| a | 1    | 0.33 | 0.5  | 0.5  |
| b | 1    | 0.33 | 0.5  | 0.5  |
| c | 0.33 | 1    | 0.66 | 0.66 |
| d | 1    | 0.33 | 0.5  | 0.5  |

### 3.3.1.1.2. La matrice des scores des similarités séquentielles

La matrice des scores des similarités séquentielles« B » est constituée de |V₁| lignes et |V₂| colonnes.

B (i, j) indique la similarité séquentielle entre deux protéines i ∈ V₁ et j ∈ V₂.

Le calcul des scores séquentiels repose sur la similarité des séquences des deux protéines de deux réseaux généré par BLAST (BLAST est un outil utilisé pour trouver les régions similaires entre les séquences des protéines en l'indiquant un score de similarité)(voir Annexe A).

Les scores séquentiels entre les protéines sont calculés en combinant entre la similarité séquentielle des deux protéines (i, j) à comparer « B⁰ »et les scores séquentiels de tous ses voisins « BN » où deux protéines sont similaires si et seulement si leurs voisins sont similaires.

B (i, j) est calculé en utilisant la formule(5) :

$$B^{t+1}(i,j) = \beta * B^0(i,j) + (1-\beta) * BN(i,j) \qquad (5)$$

Où B⁰(i, j) indique le score séquentiel de deux protéines à comparer calculé par la formule(6), et 0 ≤β≤1 un paramètre qui équilibre entre les valeurs trouvées par la formule(6) et (7):

$$B^0(i,j) = \frac{BLAST(i,j)}{max_{u \in V_1, v \in V_2}\{BLAST(u,v)\}} \qquad (6)$$

Où l'enumérateur de la formule représente la valeur de similarité séquentielle générée par l'outil BLAST, et le dénominateur indique la maximum valeur de similarité séquentielle entre toute les protéines i ∈ V₁ et j ∈ V₂.

BN (i, j) est le score séquentiel de tous les voisins de deux protéines (i, j) calculé par la formule (7) :

$$BN(i,j) = \frac{\sum_{(i',j') \in N} BLAST(i',j')}{max\{N(i)||N(j)\}} \qquad (7)$$

Pour calculer le score séquentiel, nous commençons par créer un graphe biparti $G_b(V_b, E_b)$, où $V_b$ présente deux ensembles disjoints des voisins des protéines à comparer $N(i)$ et $N(j)$ et $E_b$ présente l'ensemble des interactions entre ces voisins $(i', j') \in E_b$ où $i' \in N(i)$ et $j' \in N(j)$.

Après, on selecte un arrêt $e = (u, v)$, on ajoute « $e$ » à l'ensemble N et on répète ce processus jusqu'à parcourir tous les arrêts du graphe. Une fois l'ensemble N est déterminé, nous calculons les scores séquentiels de deux protéines à comparer « $B^0$ » et les scores séquentiels de tous les voisins de deux protéines « BN » et on les combine pour déterminer le score séquentiel final.

### 3.3.1.1.2.1. Exemple de calculer le score séquentiel entre deux protéines de G1 et G2

Etant donné deux réseaux d'IPPs G1 et G2 et les « BLAST scores » entre toute les protéines de G1 et G2 :

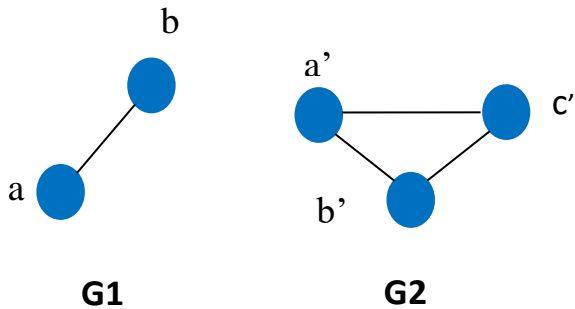

| aa' | 90.000 |
| a b' | 35.000 |
| a c' | 31.000 |
| b a' | 20.000 |
| b b' | 15.000 |
| b c' | 99.000 |

**Supposons qu'on veuille calculer B(a, b')= ?**

Le maximum nombre des voisins entre a et b'= 2

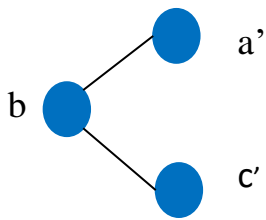

N={(b, a'), (b, c')}

$$B^0(a, b') = \frac{BLAST(a, b')}{max\ BLAST} = \frac{35.000}{99.000} = 0.35$$

$$B(a, b') = \frac{BLAST((b, a') + BLAST(b, c'))}{2} = \frac{20.000 + 99.000}{2} = 59.5$$

Supposons que $\beta = 0.999$

B (a, b')= β* B$^0$+ (1-β)*B$^N$=0.999*0.35 + (1-0.999)*59.5= 0.40

### 3.3.1.1.3. La matrice des scores des similarités fonctionnelles

La matrice des scores Fonctionnels « F » est constituée de |V$_1$| lignes et |V$_2$| colomnes, où F(i,j) indique la similarité fonctionnelle entre les protéines i ∈V$_1$ et j ∈ V$_2$ et 0≤F≤1.

Pour le calcul des scores fonctionnels, on se repose sur la méthode de Schlicker [18], basée sur la similarité sémantique entre les termes de GO de deux protéines à comparer.

GO est divisé en trois sous ontologies : BP, MF et CC. Chaque sous-ontologie est représentée par un ensemble des annotations des termes fonctionnels des protéines.

La mesure de Schlicker calcule la similarité sémantique entre les termes de sous ontologies « BP » et « MF » de deux protéines i ∈V$_1$ et j ∈V$_2$ pour déduire une liaison fonctionnelle entre eux.

Les termes du « CC » ne sont pas utilisés par la méthode de Schlicker parce que nous comparons la similarité fonctionnelle des protéines entre différentes espèces et la structure cellulaire se diffère d'une espèce à une autre.

Schlicker mesure la similarité sémantique entre les termes de deux protéines en utilisant « the relevance semantic similarity ».

Considérons deux protéines « A » et « B » avec leur annotations GO de taille respectivement « M » et « N » où GOA= {t$_1$, t$_2$, t$_3$, …, t$_m$} et GOB= {t'$_1$, t'$_2$, t'$_3$, …, t'$_n$}, une matrice « S » est calculée contenant les valeurs de la similarité sémantique entre $GO_i^A$ de protéine A et $GO_j^B$ de protéine B.

Les valeurs de la matrice sont calculées en utilisant « the relevance semantic similarity » :

S$_{ij}$= $sim_{Rel}(GO_i^A, GO_j^B)$, ∀ i∈ {1…M}, ∀j∈ {1…N}

Où

$$sim_{Rel}(t_1, t'_1) = max_{t \in set(t_1, t'_1)} \left( \frac{2.logp(t)}{logp(t_1) + logpt'_1} . (1 - p(t)) \right) \tag{8}$$

Où p(t) est la fréquence du terme « t» dans l'ensemble d'annotation.

Les lignes de la matrice « S » représentent les termes de GOA et les colonnes représentent les termes de GOB, les lignes et les colonnes représentent deux comparaisons directionnelles différentes où chaque terme de protéine « A » est comparé aux termes de protéine « B » et chaque terme de protéine « B » est comparé aux termes de protéine « A ». (Voir tableau 3.1).

|  | t'$_1$ | t'$_2$ | t'$_3$ | … | t'$_n$ | rowScore |
|---|---|---|---|---|---|---|
| t$_1$ | 0.6 | 0.4 | 0.3 | … | 0.4 | Y$_1$ |
| t$_2$ | 0.1 | 0.6 | 0.2 | … | 0.2 | Y$_2$ |
| t$_3$ | 0.3 | 0.2 | 0.6 | … | 0.7 | Y$_3$ |
| … | … | … | … | … |  | … |
| t$_m$ | 0.5 | 0.3 | 0.8 | … | 0.1 | Y$_m$ |
| columnScore | X$_1$ | X$_2$ | X$_3$ | … | Xn |  |

**Tableau3. 1: Exemple de la matrice « S ».**

La similarité sémantique est appliquée sur les deux sous ontologies« BP » et « MF » pour obtenir BPscore et Mfscore où

BPscore =max {columnScore, rowScore} et Mfscore=max {{columnScore, rowScore}

où

$$rowScore = \frac{1}{m}\sum_{i=1}^{m} \begin{array}{c} maxSij \\ 1 \leq j \leq n \end{array} \qquad (9)$$

$$colomnScore = \frac{1}{n}\sum_{j=1}^{n} \begin{array}{c} maxSij \\ 1 \leq i \leq m \end{array} \qquad (10)$$

Après avoir calculé BPscore et Mfscore en utilisant « the relevance semantic similarity », la similarité fonctionnelle entre deux protéines est calculée comme suit :

$$F = \sqrt{\frac{1}{2}\left(\frac{BPscore}{max(BPscore)}\right)^2 + \left(\frac{MFscore}{max(MFscore)}\right)^2} \tag{11}$$

Où max(BPscore) et max(Mfscore) désignent respectivement le score maximal pour « BP »et « MF».

### 3.3.1.2. La matrice des scores d'interactions

La matrice des scores d'interactions« I » est constituée de $|V_1|$ lignes et $|V_2|$ colonnes. En alignant i∈$V_1$ avec j∈$V_2$, I (i, j) indique une valeur approximative des nombres d'interactions conservées incidentent de i appartenant à G1.

Chaque nœud a une dépendance à l'un de ses voisins égale à $\frac{1}{|N(i)|}$. La dépendance indique la probabilité que l'interaction du nœud « i » avec l'un de ses voisins soient conservées.

Par exemple, si le nœud «i » a trois voisins {i', i'', i'''}, chaque interaction (i, i'), (i, i'') et (i ,i''') sera conservée avec une probabilité égale à $\frac{1}{3}$.

La sommation des dépendances des voisins de « i » indique une valeur approximative des nombres d'interactions conservées incidentent de« i » en l'alignant avec un nœud de l'autre réseau.

Par exemple si les trois voisins de i {i', i'', i'''} ont des degrés (nombre des voisins)respectivement {4, 2, 4}, alors la sommation des dépendances des voisins de «i » est égale à$\frac{1}{4} + \frac{1}{2} + \frac{1}{4} = 1$, donc en alignant i∈$V_1$ du G1 avec j∈$V_2$ du G2 la valeur attendue des nombres d'interactions conservéés incidentent de « i » est 1. Le score d'interaction entre deux protéines est calculé en utilisant la formule (12) :

$$I(i,j) = \frac{min\{\sum_{i'\in N(i)}\frac{1}{|N(i')|}, \sum_{j'\in N(j)}\frac{1}{|N(j')|}\}}{max_{K\in V_1\cup V_2}\{|N(K)|\}} \tag{12}$$

Si deux nœuds sont appariés, la valeur approximative des nombres d'interactions conservées sera changée, donc la matrice des scores d'interactions sera modifiée. Pour ce faire deux matrices sont déclarées :

### 3.3.1.2.1. La matrice d'incrémentions de nombres d'interactions conservées

La matrice « II » est une matrice de $|V_1|$ ligne et $|V_2|$ colonnes.

Soit deux protéines $i \in V_1$ et $j \in V_2$ sont appariés, $II(i, j)$ indique le nombre d'interactions conservées incidente de « i » dans G1. Supposons que deux autres protéines $u \in V_1$ et et $v \in V_2$ soient alignés où $i \in N(u)$ et $j \in N(v)$ alors l'interacation(u, i) est conservée parce qu'elle est alignée avec l'interaction(v, j) donc le nombre d'inerations conservées sera incrementé par 1.

Autrment dit, lorsque « u » et « v » sont appariés, le nombre d'inetractions conservés résultant de l'alignement de tous paire des protéines $i \in V_1$ et $j \in V_2$ où $i \in N(u)$ et $j \in N(v)$ sera incrementé par 1 et la matrice « II » sera modifiée selon la formule (13) :

$$II(i,j)^{K+1} = \begin{cases} II(i,j)^K + 1 & i \in N(u), j \in N(v) \\ II(i,j) & otherwise \end{cases} \qquad (13)$$

### 3.3.1.2.2. La matrice de suppression de dépendances

Chaque protéine $u \in V_1$ a une dépendance égale à $\frac{1}{|(N(u)|}$ à l'un de ses voisins. Après avoir aligner la protéine « u » du G1 au protéine « v » de G2, ces deux protéines seront supprimées de la matrice d'alignement donc leur dépendance doit être supprimée de la matrice des scores d'interactions de tous voisins de $i \in N(u)$ du G1 et $j \in N(v)$ du G2. Donc, pour ce faire deux matrices $C_1(i)$ et $C_2(j)$ sont créées respectivement pour G1 et G2.

Chaque valeur de $C_1(i)$ et $C_2(j)$ indique la dépendance doit être suppriméede la valeur approximative des nombre d'interations conservées de nœud « i » et « j ».

Les deux matrices $C_1(i)$ et $C_2(j)$ sont determinées selon les deux formules (14) et (15).

$$C_1(i)^{K+1} = \begin{cases} C_1(i)^K + \frac{1}{|N(u)|} & i \in N(u) \\ C_1(i)^K & otherwise \end{cases} \qquad (14)$$

$$C_2(j)^{K+1} = \begin{cases} C_2(j)^K + \frac{1}{|N(v)|} & j \in N(v) \\ C_2(j)^K & otherwise \end{cases} \quad (15)$$

Après la mise à jour des matrices $C_1(i)$ et $C_2(j)$, la matrice des scores d'interactions doit être modifiée selon la formule(16) :

$$I(i,j) = \frac{min\{d_1(i), d_2(j)\} + II(i,j)}{max_{K \in V_1 \cup V_2}\{|N(K)|\}} \quad (16)$$

Où

$$d_1(i) = \sum_{i' \in N(i)} \frac{1}{|(N(i'))|} - C_1(i) \quad (17)$$

$$d_2(j) = \sum_{j' \in N(j)} \frac{1}{|(N(j'))|} - C_2(j) \quad (18)$$

### 3.3.1.2.3. Exemple de calculer le score d'interaction entre deux protéines de G1 et G2

Etant donné deux réseaux G1 et G2 :

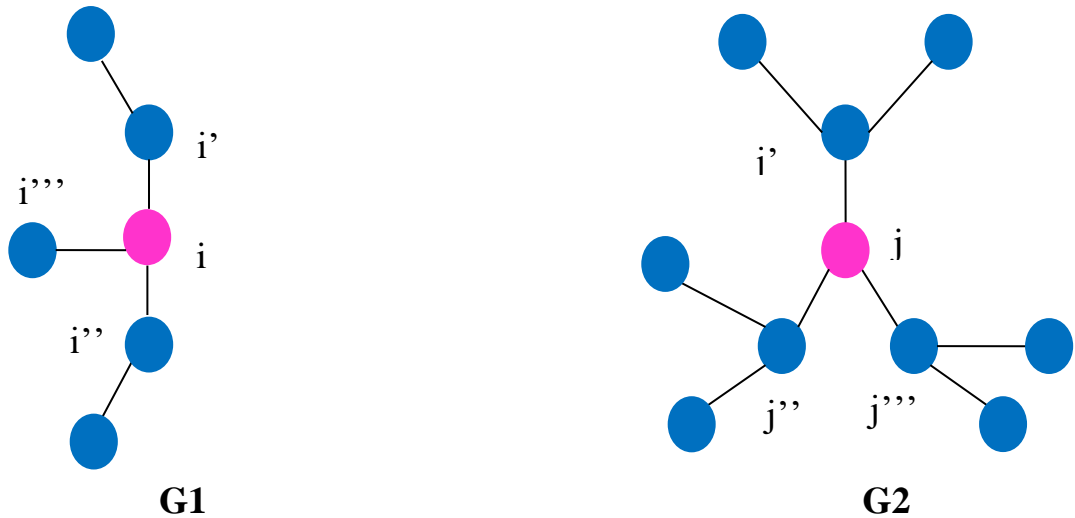

**G1**             **G2**

Suppose qu'on veuille calculer I (i, j)= ?

i à 3 voisins {i', i'', i'''} qui ont respectivement {2, 2, 1} voisins.

j à 3 voisins {j', j'', j'''} qui ont respectivement {3, 3, 3} voisins.

Chaque protéine a une dépendance à l'un de ses voisins égale à $\frac{1}{|N(i)|}$

La somme de dépendances des voisions de «i» = $\frac{1}{N(i')} + \frac{1}{N(i'')} + \frac{1}{N(i''')} = \frac{1}{2} + \frac{1}{2} + \frac{1}{1} = 2$

La somme de dépendances des voisions de « j » = $\frac{1}{N(j')} + \frac{1}{N(j'')} + \frac{1}{N(j''')} = \frac{1}{3} + \frac{1}{3} + \frac{1}{3} = 1$

Le maximum nombre des voisins entre tous les protéines de G1 et G2 est 3.

Donc I (i, j)= $\frac{min(2, 1)}{3} = 0.33$

Après la construction de la matrice des scores de similarité S (i, j) et la matrice des scores d'interactions I (i, j), la matrice des scores d'alignements est calculée selon l'équation (19).

$$A(i,j) = \lambda\, S(i, j) + (1 - \lambda) * I(i, j) \tag{19}$$

Où 0≤λ≤1 est un paramètre qui équilibre entre les valeurs de deux matrices.

### 3.3.2. La deuxième étape : l'alignement global final de deux réseaux G1 et G2

La deuxième étape consiste à trouver le meilleur alignement global entre les protéines de deux réseaux d'entrée G1 et G2. L'algorithme commence par chercher les paires des protéines avec le maximum score d'alignement et les alignées. Ensuite, la matrice des scores d'interactions est mise à jour, et par conséquent, la matrice des scores d'alignement sera modifiée basant sur les nouvelles valeurs de I (i, j). Finalement, les deux protéines alignées seront supprimées de la matrice d'alignement pour éviter de les aligner une autre fois.

Ce processus sera répété jusqu'à ce que toutes les protéines du réseau G1 soient alignées sur les nœuds de G2.

**Algorithm   PPINA (G1, G2, α, option)**

**Require:** |V$_1$|≤|V$_2$|

Construct topological score matrix T.

Given functional data, construct functional score matrix B.

Reading biological data, construct biological score matrix B.

if (option= 1)

  Computing similarity scores using topological and functional scores.

  **for** all i∈V$_1$ **do**

  for all j∈V$_2$ **do**

  S(i, j)←αT(i, j) +(1-α)*F(i, j)

  end for

  end for

  **else if**  (option= 2)

  Computing similarity scores using topological and biological scores.

  **for** all i∈V$_1$ **do**

  for all j∈V$_2$ **do**

  S(i, j)←αT(i, j) +(1-α)*B(i, j)

  **end for**

  **end for**

  **else**

  Computing similarity scores using  functional and biological scores.

```
        for all i∈V₁ do

        for all j∈V₂ do

        S(i, j) ← αB(i, j) + (1-α)*F(i, j)

        end for

        end for

        Construct Interaction score matrix I

        Given S and I, construct Alignment score matrix

        While there is an unaligned node ∈ V₁ do

        Find an analigned nodes i∈V₁ and j∈V₂ with maximum A(i, j)

        f(i) ← j

        update Interaction scores matrix

        Given S and I, update Alignement score matrix

        end while

        return f
```

**Figure3. 3: Les étapes générales de l'algorithme « PPINA ».**

## 3.4. Évaluation des résultats

Pour évaluer la qualité d'alignement de notre approche, nous testons « PPINA » sur de réseaux d'IPPs réelles.

Les fichiers d'entrée et les sorties de « PPINA » :

**Les entrées :**

- Deux réseaux IPPs (voir Annexes B) : chaque ligne dans le réseau correspond à une interaction entre deux protéines du même réseau séparé par un espace de tabulation.
    **Pour la similarité biologique** :

- Le fichier « BLAST scores » fourni par BLAST (voir Annexes A) : il faut comparer le réseau ayant le moins nombre des protéines sur le réseau ayant le plus nombre des protéines, et la troisième colonne du fichier correspond à la valeur de la similarité entre eux séparés par un espace de tabulation.

    **Pour la similarité fonctionnelle** :

- Le fichier de Gene Ontology(GO).
- Le fichier des associations des gènes.
- Le fichier des annotations des gènes.

**La sortie :**

- Un fichier contenant l'alignement global des deux réseaux d'entrée (voir Annexes A).

Nous avons évalué la qualité topologique des résultats d'alignement en utilisant les mesures d'évaluations topologiques que nous avons définies dans le chapitre 1.

L'évaluation de la qualité d'alignement topologiquement est faite par des métriques d'évaluation déjà implémenté dans l'approche NETAL [12].

### 3.4.1. Les réseaux d'IPPs

Nous avons utilisé quatre espèces eucaryotes pour évaluation tels que « Mus musculus (Mouse) », « Drosophilamelanogaster (Fly) », « Caenorhabditiselegans (Worm) »et « Rattusnorvegicus »(Rat). Les quatre réseaux d'IPP eucaryotes ont tous été téléchargés de [15]. Le nombre de protéines et les interactions de ces réseaux PPI sont indiqués dans le tableau 3.2.

**Tableau3. 2: Caractéristiques de réseaux d'IPPs de différentes espèces.**

| Espèce | nœuds | arrêts |
|---|---|---|
| Musmusculus (Mouse) | 2793 | 4085 |
| Rattusnorvegicus (Rat) | 1135 | 1290 |
| Caenorhabditiselegans (Worm) | 4305 | 7747 |
| Drosophilamelanogaster(Fly) | 7747 | 1290 |

### 3.4.2. Résultats et discussion

Dans cette section, l'algorithme « PPINA » est comparé avec d'autres algorithmes d'alignement des réseaux d'IPPs tels que, NETAL, MAPPIN et MAGNA++.

La raison de choisir ces algorithmes est leur disponibilité et la précision de leurs résultats par rapport à d'autres outils existants.

Pour les tests de « PPINA », nous avons fixé les valeurs des paramètres α = 0.0001, β= 0.999 et λ = 0.0001.

Nous avons exécuté MAPPIN, MAGNA++ et NETAL selon leurs paramètres par défaut mentionnés respectivement dans [15] [33] [12].

Nous avons validé les résultats de l'alignement de MAPPIN seulement en utilisant les métriques ME, MNE car les métriques d'évaluation topologique n'ont pas été implémentées dans cet algorithme.

Nous avons évalué la performance de « PPINA » sous plusieurs versions :

- Seulement en utilisant les scores fonctionnels « PPINA_F ».
- Seulement en utilisant les scores séquentiels « PPINA_Seq ».
- En combinant entre la matrice topologique et fonctionnelle «PPINA_TF ».
- En combinant entre la matrice topologique et séquentielle «PPINA_TS».
- En combinat entre la matrice séquentielle et fonctionnelle « PPINA_SF ».

**Tableau3. 3 : Comparaison des résultats de notre algorithme avec celui de NETAL, MPPIN et MAGNA++ pour les deux espèces « Rat-Mouse » et « Worm-Fly ». Les meilleurs résultats sont mentionnés en gras pour chaque colonne.**

| Rat-Mouse | | | |
|---|---|---|---|
| | EC | ME | MNE |
| NETAL | 0.863566 | 0.494 | 0.221 |
| PPINA_F | 0.868605 | 0.412 | 0.186 |
| PPINA_Seq | **0.872868** | 0.445 | 0.203 |
| PPINA_TF | 0.870155 | 0.398 | **0.181** |
| PPINA_TS | 0.870155 | 0.423 | 0.192 |
| PPINA_SF | 0.868992 | **0.397** | **0.181** |
| MAPPIN | - | 1 | 0.417 |
| MAGNA++ | 0.870313 | 0.805 | 0.382 |
| Worm-Fly | | | |
| | EC | ME | MNE |
| NETAL | 0.584872 | **0** | **0** |
| PPINA_F | 0.592875 | **0** | **0** |
| PPINA_Seq | **0.594811** | **0** | **0** |
| PPINA_TF | 0.593778 | **0** | **0** |
| PPINA_TS | 0.591713 | **0** | **0** |
| PPINA_SF | 0.592616 | **0** | **0** |
| MAPPIN | - | 0.19 | 0.118 |
| MAGNA++ | 0.382212 | **0** | **0** |

Selon les résultats de l'alignement de « Mouse-Rat » et « Worm-Fly » qui se trouvent dans le tableau3.3, La meilleure valeur d' « EC », « MNE » et « ME » sont trouvées par PPINA. Donc, cela indique que PPINA est capable de trouver des interactions plus conservées par rapport aux autres algorithmes, les protéines alignées par notre algorithme partagent les mêmes fonctions moléculaires et que notre algorithme aligne efficacement les protéines homologues.

### 3.5. Conclusion

Dans ce chapitre, nous avons présenté notre algorithme «PPINA » tout en détaillant les différentes étapes de notre processus d'alignement.

De plus, nous avons évalué les résultats obtenus par notre approche topologiquement et fonctionnellement en les comparants avec celui de « NETAL », « MAPPIN » et « MAGNA++». Les résultats ont montré que nos résultats ont surpassé les résultats des autres méthodes.

# Conclusion générale

Après la première phase, au cours de laquelle la plupart des efforts ont été consacrés à la découverte des interactions entre les protéines, le développement des outils et des algorithmes des calculs capables de bien analyser ces données est devenu un défi pour la découverte de nouvelles connaissances dans la recherche biologique.

L'alignement de réseaux biologique peut être utilisé pour guider le transfert des connaissances moléculaires entre des régions conservées d'espèces différentes.

Dans ce mémoire, nous avons présenté notre outil d'alignement de réseaux d'IPPs appelé « PPINA » basés sur les similarités topologiques, séquentielles et fonctionnelles des protéines.

Nous avons testé notre algorithme sur de réseaux d'IPPs réels. Les résultats montrent que PPINA a surpassé d'autres algorithmes d'alignement, et il fournit des résultats biologiquement significatifs.

**Perspectives**

L'alignement de réseaux d'IPPs a été étudié de manière approfondie au cours de ces dernières années. La qualité de l'alignement ne dépend pas seulement des informations biologiques utilisées, mais aussi de la méthodologie utilisée durant le processus d'alignement. Nos futurs travaux seront l'amélioration de la qualité d'alignement de notre approche tout en essayant à le transformer à un aligneur de multiples réseaux d'IPPs. Un autre point remarquable pour les travaux futurs est l'amélioration des mesures d'évaluations existantes.

# Annexes A

| | |
|---|---|
| A1BN54 | Q62108 |
| A1L347 | Q62108 |
| A2A432 | P20263 |
| A2A5Z6 | P51141 |
| A2A5Z6 | P63001 |
| A2A5Z6 | P70399 |
| A2A5Z6 | Q8VE28 |
| A2A5Z6 | Q9Z101 |
| A2A817 | Q62108 |
| A2A8L1 | P20263 |
| A2A8L5 | Q9CQV8 |
| A2ADH4 | P84244 |
| A2AGI2 | Q62108 |
| A2AI21 | Q62108 |
| A2AKX3 | Q61188 |
| A2AKX3 | Q62315 |
| A2AKX3 | Q921E6-3 |
| A2ARV4 | P97318 |
| A2ARV4 | Q62108 |
| A2ARV4 | Q8BL58 |
| A2ARV4 | Q8K2H6 |
| A2ARV4 | Q9D6K5 |
| A2ARV4 | Q9ERE9 |
| A2ARV4 | Q9WVI9 |
| A2ARV4 | Q9Z0G0 |
| A2AWL7 | P28574 |
| A2AWL7 | Q9CQJ4 |
| A2BGG7 | Q60631 |
| A2BH40 | O35845 |
| A2BH40 | Q61466 |
| A2CG49 | P35438 |
| A2CG49 | Q01097 |
| A2CG49 | Q62108 |
| A2CG49 | Q9CQV8 |
| A3KFM7 | Q9Z0E3 |
| A3KGF7 | Q9CXF4 |
| A7DTG3 | P63101 |

**Un exemple d'interaction pour chaque paire de protéine de réseau IPPs pour l'espèce« Mouse».**

| | |
|---|---|
| A0MZ67 | Q63787 |
| A1L132 | P16599 |
| A7U7N4 | Q62968 |
| B0BN20 | Q62968 |
| B0BN93 | Q4KMA2 |
| B1WC42 | Q62968 |
| B1WC80 | Q4KMA2 |
| B2GV74 | P54256-2 |
| B2RZ97 | Q4KMA2 |
| B5DEN5 | Q4KMA2 |
| B5DEP6 | Q4KMA2 |
| C0H5Y5 | P16599 |
| D3ZA28 | Q4KMA2 |
| D3ZP81 | Q4KMA2 |
| D3ZZM9 | Q62976 |
| D4A037 | Q4KMA2 |
| D4A8G3 | O54874 |
| D4A8G3 | Q7TT49 |
| D4ACQ8 | Q4KMA2 |
| D4AEH3 | Q4KMA2 |
| D4AEH9 | P18163 |
| O08618 | P19357 |
| O08651 | P19357 |
| O08730 | P18163 |
| O08772 | P19357 |
| O08815 | O08815 |
| O08838 | O08839 |
| O08838 | P21575 |
| O08839 | P19357 |
| O08839 | P21575 |
| O09139 | P33568 |
| O09171 | P06536 |
| O09171 | Q3T1L0 |
| O35094 | P19357 |
| O35115 | Q91V26 |
| O35142 | P19357 |
| O35147 | P49950 |
| O35179 | Q62910 |
| O35179 | Q62910-2 |

**Un exemple d'interaction pour chaque paire de protéine de réseau IPPs pour l'espèce« Rat ».**

| | | |
|---|---|---|
| O55147 | A1BN54 | 170.2 |
| P04466 | A1BN54 | 41.97 |
| P30427 | A1BN54 | 240.4 |
| P62161 | A1BN54 | 65.47 |
| Q5PQM3 | A1BN54 | 75.48 |
| Q63598 | A1BN54 | 48.52 |
| Q9HB97 | A1BN54 | 46.21 |
| Q9QXQ0 | A1BN54 | 1534.2 |
| Q9Z1P2 | A1BN54 | 1750.7 |
| P06761 | A1L347 | 761.9 |
| P48721 | A1L347 | 518.5 |
| P55063 | A1L347 | 1175.6 |
| P63018 | A1L347 | 994.2 |
| O88382 | A2A5Z6 | 50.45 |
| Q9JK71 | A2A5Z6 | 51.99 |
| O88767 | A2A817 | 163.3 |
| Q8K1P7 | A2A8L1 | 366.7 |
| O55005 | A2A8L5 | 81.26 |
| P13596 | A2A8L5 | 41.59 |
| P20417 | A2A8L5 | 155.6 |
| P26453 | A2A8L5 | 40.05 |
| Q05546 | A2A8L5 | 90.51 |
| Q62838 | A2A8L5 | 113.2 |
| Q63155 | A2A8L5 | 64.7 |
| Q63259 | A2A8L5 | 172.6 |
| Q6P6V8 | A2A8L5 | 41.97 |
| Q8VHZ8 | A2A8L5 | 65.08 |
| Q9JIR1 | A2A8L5 | 41.59 |
| P34926 | A2ADH4 | 43.13 |
| Q8K1P7 | A2ADH4 | 91.28 |
| P06882 | A2AGI2 | 194.5 |
| P19490 | A2AI21 | 239.2 |
| P19491 | A2AI21 | 234.2 |
| P19492 | A2AI21 | 221.5 |
| P35439 | A2AI21 | 1867.8 |
| Q00959 | A2AI21 | 295.8 |
| Q00960 | A2AI21 | 307.8 |
| Q00961 | A2AI21 | 268.1 |
| Q62645 | A2AI21 | 256.1 |
| Q9R1M7 | A2AI21 | 255.8 |

**« BLAST scores » entre « mouse » et « rat ».**

| | |
|---|---|
| Q22799 | Q12349 |
| G5ECT7 | Q03629 |
| G5EBV6 | P47142 |
| Q9N432 | Q08222 |
| Q19816 | P40536 |
| O45335 | P15367 |
| P55853 | Q12692 |
| O45599 | P53982 |
| H2L0C9 | P53825 |
| Q17902 | P40169 |
| G5EBL8 | P32806 |
| P48150 | P40162 |
| P34574 | P33399 |
| P55954 | Q03231 |
| P48154 | P38833 |
| Q1ZXT3 | P35200 |
| Q19207 | P47172 |
| Q9U2M6 | P08525 |
| P30642 | P40340 |
| Q18194 | Q99216 |
| O45436 | P04050 |
| Q18192 | Q03862 |
| G5EDQ5 | P32915 |
| Q20805 | P42951 |
| G5EDA1 | P40094 |
| O17641 | P54000 |
| Q19988 | P53251 |
| Q9XVX4 | Q07535 |
| P91249 | P43598 |
| P34766 | P09435 |
| Q93572 | P38708 |
| Q65ZH5 | P42834 |
| G5EG76 | P32263 |
| Q18529 | P14540 |
| O45087 | P87271 |
| G5EBX2 | P38304 |
| C6KRN1 | P04386 |
| Q23064 | Q12306 |
| O16474 | P39110 |

**Le résultat d'alignement global de « PPINA ».**

```
AnalyseFunctionalData()
{

A.Read(network1.name, 1);
B.Read(network2.name, 1);

nodesA = A.get_cnodes();
nodesB = B.get_cnodes();

long templ;

for(long i =0; i!= sizeA; ++i) {  templ = allnodes.Add(nodesA->GetLabel(i));
for(long i =0; i!= sizeB; ++i) {  templ = allnodes.Add(nodesB->GetLabel(i));

char ontology_file[] = "gene_ontology.1_2.obo";
char gene_association_file[] = "intact.gene_assocs.txt";
Ont.read_go_ids(ontology_file);
Ont.read_relations(ontology_file);
Ont.add_protein_associations(gene_association_file, allnodes.get_trie());
FILE *fi = fopen("mouse_rat.gene_association", "r");
allnodes.add_gene_associations_isoform(fi, &Ont);
fclose(fi);
}
```

**La fonction d'analyse les données fonctionnelles.**

```
computeFunctSimilarity(int node1, int node2)

{

double x = Ont.function_similarity_schlicker(node1, node2);

return x;

}

double function_similarity_schlicker(VLong ids1, VLong ids2)//***+
    {
double *out = (double*)calloc(2, sizeof(double));
    out[MF] = GOScore(gos1[MF], gos2[MF]);
out[BP] = GOScore(gos1[BP], gos2[BP]);
    double out = sqrt(x[MF] *x[MF] + x[BP]*x[BP])/sqrt(2);
    free(x);
```

**La fonction de calculer les scores fonctionnels.**

```cpp
AnalyseSequenceSimilarity(int node1, int node2)
{
            ifstream inputFile;
            string token1,token2,line, blastFile;
            float token3;
            int id1, id2;

blast = newfloat*[network1.size];
    bioI = newfloat*[network1.size];
tempBio = newfloat*[network1.size];

    for (int c=0; c<network1.size; c++)
      {
    blast[c]=newfloat[network2.size];
    bioI[c]= newfloat[network2.size];
tempBio[c]=newfloat[network2.size];

      }
    for (int c1=0; c1<network1.size; c1++)
      {
    for (int c2=0; c2<network2.size; c2++)
        {
    blast[c1][c2]=0;
    bioI[c1][c2]=0;
tempBio[c1][c2]=0;

        }
      }

    blastFile = network1.name;
    blastFile.append("-");
    blastFile.append(network2.name);
    blastFile.append(".sim");

    inputFile.open(blastFile.c_str());
    while (getline(inputFile, line))
     {
    istringstream tokenizer(line);

    getline(tokenizer, token1, '\t');
        id1= network1.mapName[token1];
    getline(tokenizer, token2, '\t');
        id2= network2.mapName[token2];
    tokenizer>> token3;
    if(maxScore<token3) maxScore = token3;
    tempBio[id1][id2]=token3;
     }
    for (int c1=0; c1<network1.size; c1++)
    for (int c2=0; c2<network2.size; c2++)
    blast[c1][c2] = tempBio[c1][c2];
            for (int c1=0; c1<network1.size; c1++)
                {
                    for (int c2=0; c2<network2.size; c2++)
                      {
```

**La fonction d'analyser les données séquentielles.**

```
CompNeiBioloSimilarity ()

{
    float *simN;

    float tempN = 0;

    int degN1 = network1.deg[node1];
    int degN2 = network2.deg[node2];

    if (network1.deg[node1] <= network2.deg[node2])
    {
        simN = newfloat[network1.deg[node1] ];

        for(int i=0; i<network1.deg[node1]; i++)
        {
            simN[i] = 0;
        }
        for(int i=0; i<network1.deg[node1]; i++)
        {
            for(int j=0; j<network2.deg[node2]; j++)
            {
                simN[i] += blast[network1.neighbor[node1][i] ][ network2.neighbor[node2][j] ];
            }
            tempN += simN[i];
        }
    }
    else
    {
        simN = newfloat[network2.deg[node2] ];
        for(int i=0; i<network2.deg[node2]; i++)
        {
            simN[i] = 0;
        }
        for(int i=0; i<network2.deg[node2]; i++)
        {
            for(int j=0; j<network1.deg[node1]; j++)
            {
                simN[i] += blast[network1.neighbor[node1][j] ][ network2.neighbor[node2][i] ];
            }
            tempN += simN[i];
        }
    }
    if( degN2 >= degN1 && degN2 > 0 )
        tempN /= degN2;
    elseif( degN1 > degN2 && degN1 > 0)
        tempN /= degN1;
    else
        tempN = 0.0;
```

**La fonction de calculer les scores biologiques de tous les voisins de deux nœuds à aligner.**

```
ComputeBioloSimilarity(int node1, int node2)
{

float B= ((1-bb)*compNeiBioloSimilarity(node1, node2)+bb*(biol[node1][node2]));
return B;
}
```

**La fonction de calculer les scores biologiques.**

# Bibliographies

# Wébographies